\documentstyle[12pt,epsfig]{article}

\def\be{\begin{equation}}
\def\ee{\end{equation}}

\begin{document}

\begin{titlepage}
\setlength{\textwidth}{5.0in}
\setlength{\textheight}{7.5in}
\setlength{\parskip}{0.0in}
\setlength{\baselineskip}{18.2pt}
\setlength{\footskip}{0.5in}
\setlength{\footheight}{0in}

\renewcommand{\thefootnote}{\fnsymbol{footnote}}


\vspace{0.3cm}

\begin{center}
{\Large\bf Local free-fall temperature \\
of a RN-AdS black hole}
\end{center}

\begin{center}
{Yong-Wan
Kim\footnote{Electronic address: ywkim65@gmail.com}$^{1}$,
Jaedong
Choi\footnote{Electronic address: choijdong@yahoo.co.kr}$^{2}$}
and Young-Jai
Park$\footnote{Electronic address: yjpark@.sogang.ac.kr}^{3}$,\par
\end{center}

\begin{center}
{${}^{1}$Institute of Basic Science and School of Computer Aided
Science,
\\Inje University, Gimhae 621-749, Korea, \\}
{${}^{2}$Center of Defense Management, Korea Institute for Defense Analysis,}\par
{39 Hoegiro Dongdaemun Seoul 130-010, Korea,}\par
{and }\par
{${}^{3}$Department of Physics and BK21 Program Division, \\
Sogang University, Seoul 121-742, Korea, \\
Service Systems Management and Engineering and WCU Program Division, \\
Graduate School of Business, Sogang University, Seoul 121-742, Korea}\par
\end{center}

\vskip 0.5cm
\begin{center}
{\today}
\end{center}

\vfill

\begin{center}
{\bf ABSTRACT}
\end{center}
We use the global embedding  Minkowski space (GEMS) geometries of
a (3+1)-dimensional curved Reissner-Nordstr\"om(RN)-AdS black hole spacetime
into a (5+2)-dimensional flat spacetime to define
a proper local temperature, which remains finite at the event horizon, for freely falling observers
outside a static black hole. Our extended results include the known limiting cases of
the RN, Schwarzschild--AdS, and Schwarzschild black holes.

\vskip 0.5cm
\noindent
PACS number(s): 04.70.Dy, 04.20.Jb, 04.62.+v
\end{titlepage}

\newpage

\section{Introduction}

Unruh \cite{unr,davies75} proposed that a uniformly accelerated
observer with proper acceleration $a$ in flat spacetime will
detect thermal radiation at the so-called Unruh temperature, $T_U
= \frac{a}{2\pi}$. After his work, it has been known that a
thermal Hawking effect on a curved manifold \cite{hawking} can be
looked at as an Unruh effect in a higher dimensional flat
space-time. Non-trivial works of isometric embeddings of the RN
\cite{des-prd}, RN-AdS \cite{ksp}, Kerr \cite{ghr} black holes and
M2-, D3-, M5-branes \cite{gibbons} into flat spaces have been
studied to get some insight of the global aspect of the spacetime
geometries. Moreover, several authors \cite{des,des98,bec,gon}
have also shown that global embedding Minkowski space (GEMS)
approach
\cite{kasner,fro,friedman,goenner,ros,n,cg,hongkp,hongkp00,hongkp01}
of which a hyperboloid in a higher dimensional space corresponds
to original curved space could provide a unified derivation of
temperature for a wide variety of curved spaces. Still, the GEMS
approach has been used to understand the relation between the
thermal Hawking temperature and the Unruh Effect
\cite{rw,CTGS04,SDL04,Tian05,Radu06,Langlois06,RT08}.

Recently, Brynjolfsson and Thorlacius have used the global
embedding of Schwarzschild(-AdS), and RN black holes spacetime
into higher dimensional flat spacetimes to define a local
temperature for  freely falling observers outside a static black
hole, separately \cite{bt}. In fact, they have successfully
defined a local temperature by using the fact that there are
special turning points of radial geodesics where freely falling
observers are momentarily at rest with respect to black hole. As a
result, they have shown that the local free-fall temperature
remains finite at the event horizon, while it approaches the
Hawking temperature in asymptotically flat spacetime. Moreover,
they have also shown that freely falling observers outside an AdS
black hole do not see any high-temperature thermal radiation even
if the Hawking temperature of such black holes can be arbitrarily
high. Very recently, Greenwood and Stojkovic have also shown for
the case of the Schwarzschild black hole that the temperature is
finite at the horizon in Eddington-Finkelstein reference frame,
where the observer is not accelerated \cite{gs}.

On the other hand, ever since the discovery that thermodynamic
properties of black holes in anti-de Sitter (AdS) spacetime are
dual to those of a field theory in one dimension fewer, there has
been of much interest in RN-AdS black hole \cite{rnads}, which now
becomes a prototype example
\cite{hp,pel,Mitra99,HR99,WAS00,hcp,myu,MKP081,MKP082,mkp1} to study
this AdS/CFT correspondence \cite{wit}.

In this paper we will use the global embedding of the RN-AdS black
hole spacetime into a (5+2) dimensional flat spacetime \cite{ksp} to define a
desired local temperature for observers in radial free fall
outside a static black hole as a generalization of the previous
work \cite{bt}. As a result, this generalization includes the known limiting
cases \cite{bt} of the Schwarzschild-AdS (SAdS) and RN black holes
in (5+2)-dimensions, and Schwarzschild black hole in
(5+1)-dimensions through the successive truncations. In Sec. 2, we
briefly recapitulate the structure of the RN-AdS black holes,
which are classified by the charge $Q$ with a negative
cosmological constant $\Lambda$. In Sec. 3, we first shortly
review the $(3+1)$ dimensional RN-AdS embedding into the (5+2)
dimensional flat space, which had been obtained by two of us
\cite{ksp}, and use this GEMS method to define a proper local
temperature for observers in free fall outside a static black
hole. In Sec. 4, we also show that our results in the GEMS of the
RN-AdS space systematically include those of the known limiting
GEMS geometries, which are the RN, SAdS, and Schwarzschild,
through the successive truncation procedure of parameters in the
original curved space.
Finally, we present summary in Sec. 5.

\section{Structure of RN-AdS black hole}

Let us consider the line element of the $(3+1)$ dimensional RN-AdS
spacetime
with a negative cosmological constant $\Lambda = -3/l^2$
\cite{rnads} as
\begin{equation} \label{metric}
 ds_4^2= -f(r,m,Q,l)dt^2 + f^{-1}(r,m,Q,l) dr^2
         +r^2(d\theta^2+\sin^2\theta d\phi^2),
\end{equation}
where $f(r,m,Q,l)$ is given by
\be
f(r,m,Q,l) =
1-\frac{2m}{r}+\frac{Q^{2}}{r^{2}} + \frac{r^{2}}{l^{2}}.
\ee
Here, $m$ and $Q$ are the black hole mass and charge, respectively.

This space-time is asymptotically described by the AdS. Then, the
inner ($r_-$) and the outer ($r_+ $) horizons are obtained from the
condition of  $f(r_\pm,m,Q,l)=0$~\cite{bc}. The Arnowitt-Deser-Misner (ADM)
mass of the RN-AdS black hole and its event horizon radius $r=r_+$
are related as
\begin{equation} \label{mass0}
m = \frac{1}{2} \left[r_+ +\frac{r^3_+}{l^2}+\frac{Q^2}{r_+}\right].
\end{equation}
In this work we consider the case of fixed-charge ensemble~\cite{CEJM} with $Q < Q_c$
where $Q_c$ will be determined in Eq. (\ref{crit}).

\begin{figure*}[t!]
   \centering
   \includegraphics{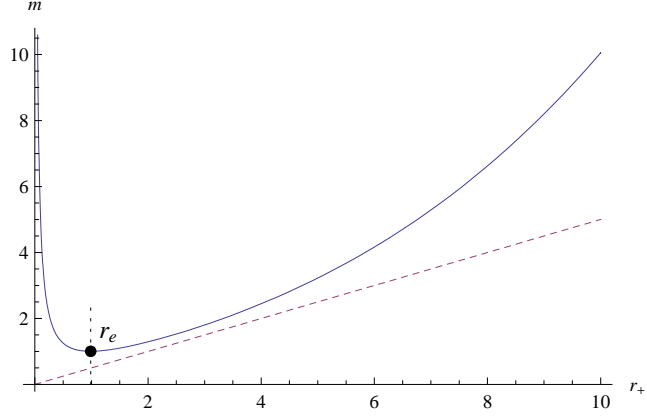}
\caption{Mass function for the RN-AdS black hole with $Q=1<Q_c=l/6$
and $l=10$. Dashed line is for the Schwarzschild black hole. The
extremal radius $r_e$ is denoted by the symbol $\bullet$.}
\label{fig.1}
\end{figure*}

The surface gravity $k_H$ is given by
\begin{equation} \label{sgrna}
 k_H(r_+,Q,l) \equiv \frac{1}{2} \left.\frac{df}{dr}\right|_{r=r_+} = \frac{(r_+^2-Q^2)l^2 +3r_+^4}{2r_+^3l^2}.
\end{equation}
Then, the Hawking temperature $T_H$,
which is the temperature of the radiation
as measured by asymptotic observer, is given by
\begin{equation} \label{aas}
T_{H}(r_+,Q,l)= \frac{k_H}{2\pi} = \frac{1}{4\pi}\left(
\frac{1}{r_+}-\frac{Q^2}{r_+^3}+\frac{3r_+}{l^2}\right).
\end{equation}
Note that the Hawking temperature of large AdS black holes grows
linearly with $r_+$, and becomes arbitrarily high for very large
black holes. However, as we will see below, this does not mean
that the physical temperature measured by an observer in free fall
becomes large outside large AdS black holes \cite{bt}.

On the other hand, the local fiducial temperature, which is the temperature of the radiation
as measured by a fiducial observer, is given by
\begin{equation}
T_{FID}(r) = \frac{T_H}{\sqrt{f(r,m,Q,l)}}.
\end{equation}
Here, the fiducial observer means an observer who remains at rest
with respect to black hole at a fixed distance. Note that the
fiducial temperature $T_{FID}$ diverges at the black hole event
horizon. On the other hand, in asymptotically flat spacetime
$T_{FID}$ approaches the Hawking temperature asymptotically far
away from the black hole, while in asymptotically AdS
spacetime $T_{FID}$ goes to zero far away from the black hole.

Moreover, using the outer and inner horizons, the mass, charge, and Hawking temperature are expressed as follows
\begin{eqnarray}\label{mqt}
 m(r_+,r_-) &=&\frac{1}{2}\left[r_++r_-+
       \frac{r_+^4-r_-^4}{l^2(r_+-r_-)}\right], \nonumber \\
 Q^2(r_+,r_-) &=&
        r_+r_-\left[1+\frac{r_+^3-r_-^3}{l^2(r_+-r_-)}\right], \nonumber \\
 T_{H}(r_+,r_-) &=& \frac{r_+ - r_-}{4\pi l^2 r^2_+}\left(l^2 + 3r^2_+ + 2r_+ r_- + r^2_- \right),
\end{eqnarray}
respectively. For the degenerate case ($r_+=r_-=r_e$), we
have an extremal black hole with $m=m_e=r_e$ as shown in Fig.1. In
general, one has an inequality of $m > m_e$. Then, using the Eqs.
(\ref{mass0}) and (\ref{aas}), the heat capacity
$C_Q=(dm/dT_{H})_Q$ for the fixed-charge $Q$ takes the form
\begin{equation}\label{aac}
C_Q(r_+,Q,l)= 2\pi r_+^2
\left[\frac{3r_+^4+l^2(r_+^2-Q^2)}{3r_+^4-l^2(r_+^2-3Q^2)}\right].
\end{equation}

\begin{figure*}[t!]
   \centering
   \includegraphics{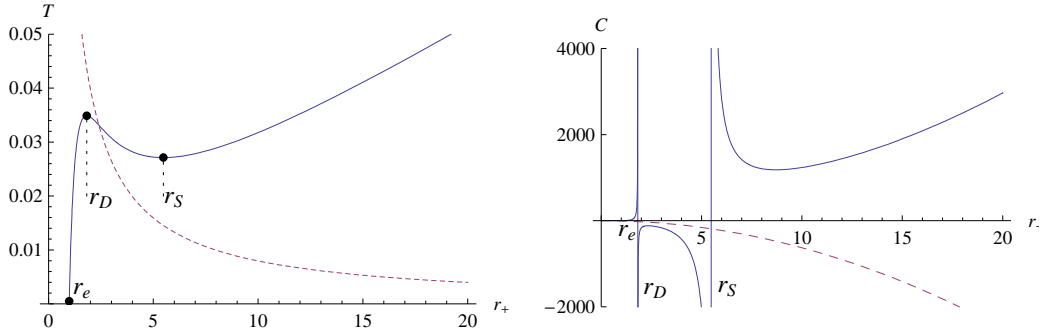}
\caption{Thermodynamic quantities of the RN-AdS black hole as
function of horizon radius $r_+$ with fixed $Q=1<Q_c$ and $l=10$:
temperature $T_H$ with $T_D=0.035,~T_s=0.027$, and heat capacity
$C$. Dashed curves are for the Schwarzschild black holes. }
\label{fig.2}
\end{figure*}

The global features of thermodynamic quantities for $Q<Q_c=l/6$ are shown in Fig.
2.  Here we observe the local minimum $T_H=T_0$ at
$r_+=r_{s}$ (SAdS-like black hole), in addition to the zero temperature
$T_H=0$ at $r_+=r_e$(extremal RN-like black hole) and the local maximum
$T_H=T_D$ at $r_+=r_{D}$ (Davies' point of RN black hole).
Therefore, it seems to be a combination of the RN and SAdS black
holes~\cite{mkp1}.

Furthermore, we observe that $C_Q=0$ and $T_H=0$ at $r_+=r_e$, where
\begin{equation}
r_e^2=\frac{l^2}{6}\left(-1+\sqrt{1+\frac{12Q^2}{l^2}}\right),
\end{equation}
and  the heat capacity  blows up at $r_+=r_D$ and $r_s$, where these
satisfy
\begin{equation}\label{crit}
r_D^2=\frac{l^2}{6}\left(1-\sqrt{1-\frac{36Q^2}{l^2}}\right),
~r_s^2=\frac{l^2}{6}\left(1+\sqrt{1-\frac{36Q^2}{l^2}}\right).
\end{equation}
These points exist only for $Q \leq Q_c = l/6$. For the $Q=Q_c$ case, we have
$r_D=r_s=l/\sqrt{6}$.
The local stability is usually determined  by the positive sign of
heat capacity by considering evaporation and absorbing processes
of a black hole~\cite{TL}. For example, the heat capacity of the
Schwarzschild black hole is $-2\pi r_+^2$, which means that this
isolated black hole is not in equilibrium in asymptotically flat
spacetimes. Based on the local stability of heat capacity, the
RN-AdS black holes of $Q<Q_c$ can be split into stable small AdS
black hole with $C_Q>0$ being in the region of $r_e<r_+<r_D$,
intermediate unstable black hole with $C_Q<0$ in the region of
$r_D<r_+<r_s$, and stable large AdS black hole with $C_Q>0$ in the
region of $r_+>r_s$.

Note that large AdS black holes with $r_+ \gg l$~\cite{ht}
correspond to high-temperature thermal states in the dual gauge
theory \cite{wit}, while small AdS black holes with $r_+ \ll l$
can be viewed as more-or-less ordinary RN black holes in a
cosmological background with a negative cosmological constant.

\section{Local Temperature of RN-AdS Black Hole\\ in the GEMS Approach}

Ten years ago, through the GEMS approach, which makes the curved
spacetime possibly embedded in a higher dimensional flat spacetime
\cite{kasner,fro,friedman,goenner,ros,n}, two of us had obtained a
$(5+2)$-dimensional isometric embedding \cite{ksp} of the RN-AdS
spacetime with metric $\eta_{IJ} = {\rm diag.}(-1, 1, 1, 1,
1,1,-1)$ $(I,J = 0,1,...,6)$ as
\begin{eqnarray}
\label{rna}
&& z^0=k_H^{-1}\sqrt{f(r,m,Q,l)}\sinh (k_Ht),\nonumber\\
&& z^1=k_H^{-1}\sqrt{f(r,m,Q,l)}\cosh (k_Ht),\nonumber\\
&& z^2=\int dr
        \left(
              \frac{Q^2l^2}{[rr_H(r^2+rr_H+r_H^2)+(rr_H-Q^2)l^2]}
         \right. \nonumber \\
&& ~~~  + \left. \frac{r^2_H l^2(r^2 + rr_H + r_H^2)
            [(r_H^2-Q^2)^2 l^4 + r_H^6 (r_H^2+2 l^2)]}
        {r^2 [3r_H^4 + (r_H^2-Q^2)l^2]^2
           [rr_H(r^2 + rr_H + r_H^2)+(rr_H-Q^2)l^2]}
              \right)^{1/2},   \nonumber \\
&& z^3=r\sin\theta\cos\phi, \nonumber\\
&& z^4=r\sin\theta\sin\phi, \nonumber\\
&& z^5=r\cos\theta, \nonumber \\
&& z^6=\int dr \left(
  \frac{Q^2 l^4 r_H^6 [4(r r_H-Q^2)l^2+10r^4
        +2rr_H(r^2+rr_H+2r^2_H)]}
       {r^4[3r_H^4 + (r_H^2-Q^2)l^2]^2
       [rr_H(r^2+rr_H+r_H^2)+(rr_H-Q^2)l^2]}
       \right.  \nonumber \\
&& ~~~+ \left.
       \frac{r r_H(r^2+rr_H+r_H^2)
        (4r_H^6 l^2+[3r_H^4+(r_H^2-Q^2)l^2]^2)}
       {[3r_H^4+(r_H^2-Q^2)l^2]^2
        [rr_H(r^2+rr_H+r_H^2)+(rr_H-Q^2)l^2]}
       \right)^{1/2}
\end{eqnarray}
with an additional spacelike $z^2$ and a timelike $z^6$
dimensions. Here, we have denoted the outer horizon $r_+$ as $r_H$. As
a result, the (3+1)-dimensional curved spacetime of the RN-AdS is
seen as the hyperboloid embedded in the (5+2)-dimensional flat
spacetime. It would be easily verified inversely that the
following flat metric in the (5+2)-dimensional space defined as
the coordinates (\ref{rna}) gives the original RN-AdS metric
(\ref{metric}) correctly
\begin{eqnarray}
ds_7^2 \equiv  \eta_{IJ} dz^I dz^J = ds_4^2.
\end{eqnarray}
This equivalence between the (5+2)-dimensional flat embedding
spacetime and original (3+1)-dimensional curved one is
well-established for the definition of isometric embedding,
mathematically developed by several authors
\cite{gibbons,nash,nash56,clake70}.

Now, let us consider a freely falling observer dropped from rest
at $\tau = 0$ and $r = r_0$. Note that this observer differs from
a fiducial observer who remains at rest with respect to the black hole at
a fixed distance. The equations for the orbit are
\begin{eqnarray}
\label{orb1}
\frac{dt}{d\tau} &=& \frac{\sqrt{f(r_0,m,Q,l)}}{f(r,m,Q,l)}, \\
\label{orb2} \frac{dr}{d\tau} &=& \sqrt{f(r_0,m,Q,l)-f(r,m,Q,l)}.
\end{eqnarray}
Then, the 7-acceleration $a^I_7$ definded in the
$(5+2)$-dimensional embedded flat spacetime is spacelike for all
timelike orbits, and at the turning point $r = r_0$, where the
observer is dropped from rest, its squared magnitude $a^2_7 (= \eta_{IJ} a^I_7 a^J_7)$ is given by
\begin{eqnarray}
 a^2_7&=&\left[4(1+x)-(c^2+1)(c^2+5)x^2-(c^2+1)^2x^3(1+x+x^2)\right.\nonumber\\
   && -b^2(1+b+b^2+c^2)^2x^2 (1+x+x^2+x^3+4x^5)\nonumber\\
   && +\left.2b(1+b+b^2+c^2)x^2(2+4x+(c^2+1)(1+x+x^2+x^3+2x^4))\right]\nonumber\\
  &&/\left[4 c^2  r_H^2 (b x-1) \left(1+(1+b)
              x+\left(1+b+b^2+c^2\right) x^2\right)\right],
\end{eqnarray}
where $x \equiv r_H /r$, $Q^2 \equiv bl^2 (1+b+b^2 +c^2)/c^4$,
$b\equiv r_-/r_H$, and $c \equiv l /r_H$.

Taking the local temperature $T_{FFAR}=a_7/2\pi$ measured by the
freely falling observer at rest to be the local Unruh temperature
of the corresponding observer in the (5+2)-dimensional flat
spacetime, we obtain
\begin{eqnarray}
\label{temp07} T^2_{FFAR} &=&
\left[4(1+x)-(c^2+1)(c^2+5)x^2-(c^2+1)^2x^3(1+x+x^2)\right.\nonumber\\
   && -b^2(1+b+b^2+c^2)^2x^2 (1+x+x^2+x^3+4x^5)\nonumber\\
   && +\left.2b(1+b+b^2+c^2)x^2(2+4x+(c^2+1)(1+x+x^2+x^3+2x^4))\right]\nonumber\\
  &&/\left[16\pi^2 c^2  r_H^2 (b x-1) \left(1+(1+b)
              x+\left(1+b+b^2+c^2\right) x^2\right)\right].
\end{eqnarray}
In the limit of $c\rightarrow\infty$ (or, $l\rightarrow\infty$),
the temperature is reduced to the isometrically flat embedded one
for the RN
\begin{eqnarray} \label{temp7r1}
 T^2_{FFAR} &=& \frac{(1 -b)^2 (1 + x + x^2 + x^3)-
4 bx^4 + 4 b^2 x^5}{16\pi^2 r_+^2  (1 - b x)}.
\end{eqnarray}
In the limit of $b\rightarrow 0$ (or, $Q\rightarrow 0$), the
temperature becomes the corresponding flat embedded one for the
SAdS
\begin{eqnarray} \label{temp7s}
 T^2_{FFAR} &=& \frac{- 4(1+x) + (c^2+1) (c^2+5) x^2 + (c^2+1)^2 x^3 (1 + x + x^2)}
                   {16\pi^2 c^2 r_H^2 [1 + x + (c^2+1)
                   x^2]}.\nonumber\\
\end{eqnarray}
These limits are exactly the same with the ones obtained in
\cite{bt}.

The free-fall temperature (\ref{temp07}) can be further simplified in
the two interesting limits: at spacial infinity $r \rightarrow
\infty$~($x \rightarrow 0$) and  at event horizon $r \rightarrow
r_H$~($x \rightarrow 1$). At spacial infinity $r \rightarrow
\infty$, one obtains imaginary temperature as
\begin{equation}\label{sadsT}
T^2_{FFAR} \rightarrow - \frac{1}{4\pi^2 l^2},
\end{equation}
which is allowed for a geodesic observer who follows a spacelike
motion in empty AdS space \cite{des}.  Note also that from $l=c~r_H$
as $c$ increases $T^2_{FFAR}$ becomes negatively small, while as $c$
decreases $T^2_{FFAR}$ becomes negatively large, however, not
possibly positive. On the other hand, at event horizon, $r
\rightarrow r_H$, one obtains
\begin{equation}
 T^2_{FFAR} \rightarrow \frac{(1-2b)c^2-2b(1+b+b^2)}{4\pi^2 l^2},
\end{equation}
which has always imaginary temperature when $b\ge 1/2$. However,
when $0<b<1/2$, the free fall temperature $T^2_{FFAR}$ turns out to
be real and positive if the following condition is satisfied
\begin{equation}\label{cond1}
c\ge \sqrt{\frac{2b(1+b+b^2)}{1-2b}}.
\end{equation}

\begin{figure*}[t!]
   \centering
   \includegraphics{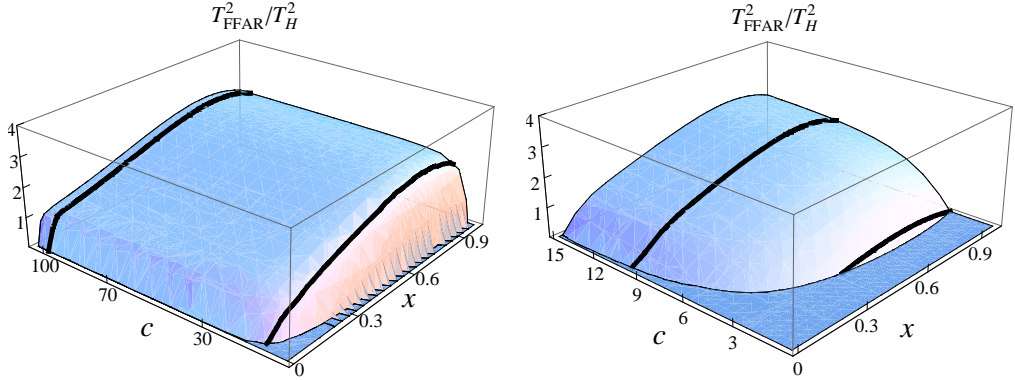}
\caption{Free-fall temperature $T^2_{FFAR}$ in units of $T^2_H$ for
the $b=0.4$ case. The solid curves are for $c=100, 10, 2.5$
describing the permitted radiation range $x$.}
 \label{fig.3}
\end{figure*}
\begin{figure*}[t!]
   \centering
   \includegraphics{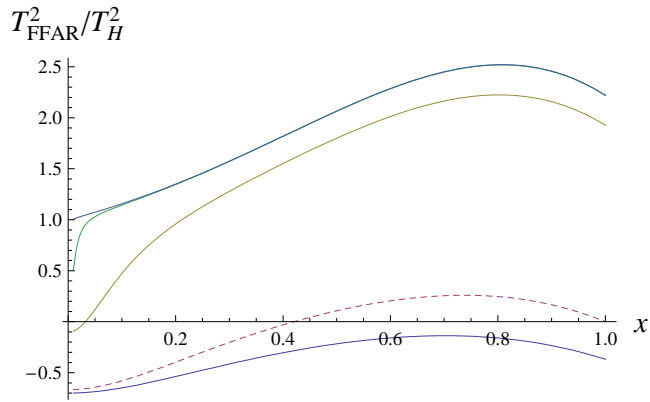}
\caption{Slices of free-fall temperature $T^2_{FFAR}$ for the
$b=0.4$ case. The curves are for $c=1000, 100, 10, 2.5, 1.9$ from
top to down. The curve for $c=1000$ actually represents the
Schwarzschild limit, and the dashed curve describes that the free
fall temperature $T^2_{FFAR}$ vanishes at event horizon $x=1$.}
 \label{fig.4}
\end{figure*}

Now, let us look into their general features in detail for the extended case of
the RN-AdS black holes, which are described by the parameters $b$ and $c$. First, Fig.
\ref{fig.3} shows the free fall temperature $T^2_{FFAR}$ of the
RN-AdS black holes, seen by the freely falling observer from the
rest, in units of the Hawking temperature $T^2_H$ given by
\begin{equation}\label{Htemp07}
T_H=\frac{(1-b)(3+2b+b^2+c^2)}{4 \pi c^2r_H}.
\end{equation}
On the other hand, Fig. \ref{fig.4} shows the ratio of the free fall
temperature $T^2_{FFAR}$ to the Hawking temperature $T^2_H$ from the
horizon ($x=1$) to the spacial infinity ($x=0$) according to the
different values of $c$. In particular, for the $b=0.4$ case,
representing one of the RN-AdS black holes, two bold curves of left
panel in Fig. \ref{fig.3} are for $c = 100, 10$, which correspond to
small black holes. On the other hand, the curves of right panel in
Fig. \ref{fig.3} are for $c=10, 2.5$, where the last $c=2.5$ case is
part of large black holes. The curve of $c=1.9$ is not shown in Fig.
\ref{fig.3}, but shown in Fig. \ref{fig.4} because it is not
satisfied Eq. (\ref{cond1}) and thus has always negative for all
$x$. This reflects the fact of no thermal radiation. In Fig.
\ref{fig.4}, the dashed curve is for $c=2.5$ in which value the free
fall temperature $T^2_{FFAR}$ at event horizon vanishes. When
$c>2.5$, the temperatures $T^2_{FFAR}$ are all real and finite as
shown in Fig. \ref{fig.4} at event horizon. The curve for $c=1000$
in Fig. \ref{fig.4} represents the Schwarzschild limit.

\begin{figure*}[t!]
   \centering
   \includegraphics{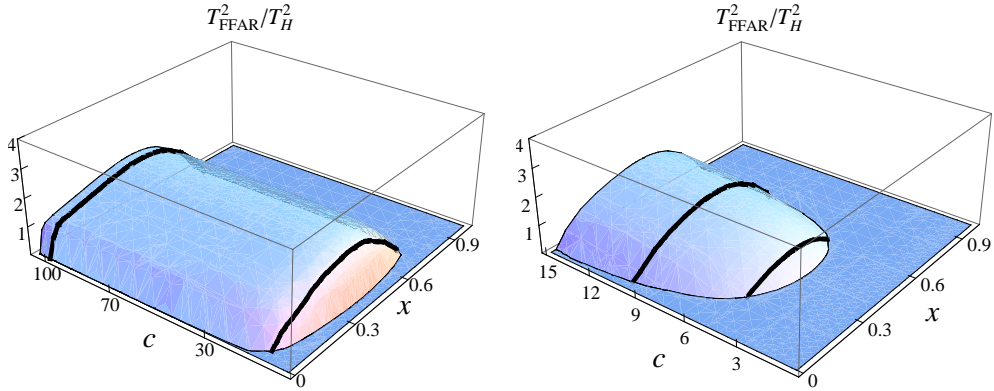}
\caption{Free-fall temperature $T^2_{FFAR}$ for the $b=0.6$ case.
The solid curves are for $c=100, 10, 5$.} \label{fig.5}
\end{figure*}

\begin{figure*}[t!]
   \centering
   \includegraphics{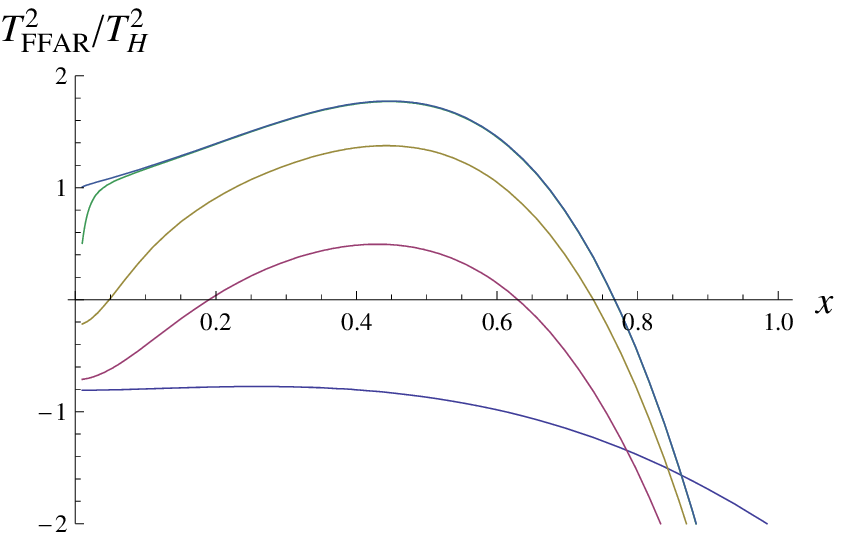}
\caption{Slices of free-fall temperature $T^2_{FFAR}$ for the $b=0.6$ case.
The curves are for the $c=1000, 100, 10, 5, 1$ cases from top to down.}
\label{fig.6}
\end{figure*}

Figs. \ref{fig.5} and \ref{fig.6} describe the RN-AdS black holes
for the $b=0.6$ case. Two bold curves of left panel in Fig.
\ref{fig.5} describe the squared free-fall temperatures for $c =
100, 10$, and the last curve of right panel is for $c=5$. Compared
with the RN-AdS case with $b=0.4$, all these local temperatures
become negative for freely falling observer before arriving at the
event horizon showing that the observer feels no thermal radiation
near the horizon.
\begin{figure*}[t!]
   \centering
   \includegraphics{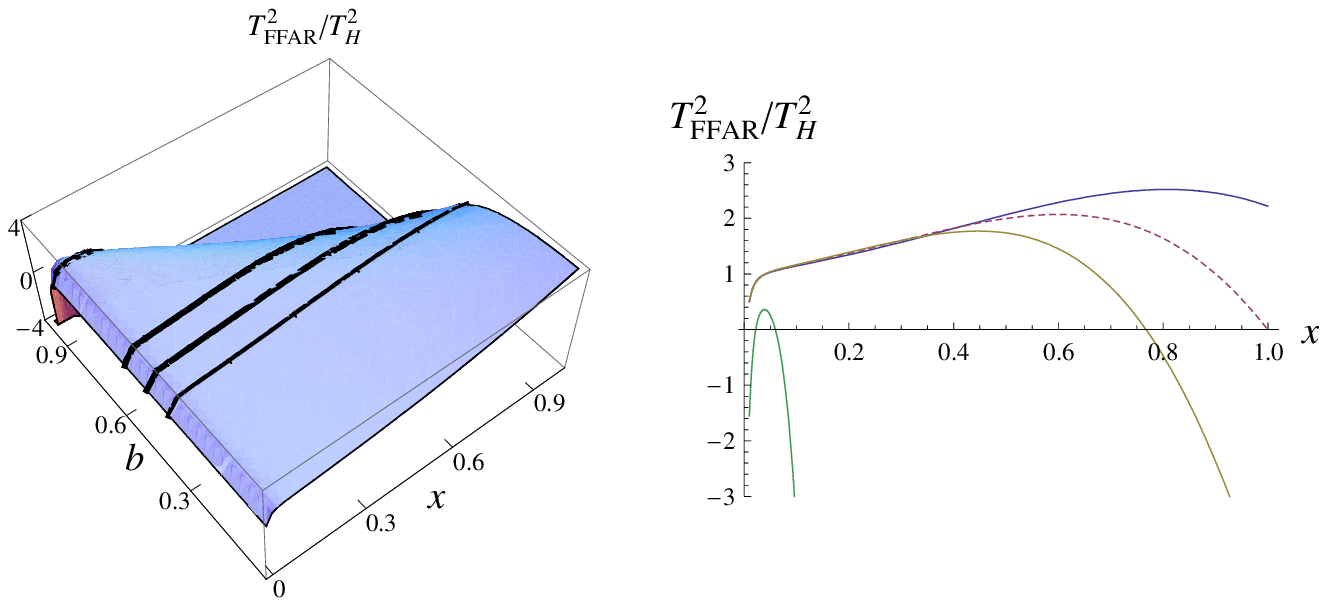}
\caption{Left panel: Free -fall temperature $T^2_{FFAR}$ for the
RN-AdS black holes with $c=100$. The solid curves are for the $b=0.4,
0.5, 0.6, 1$ cases. Right panel: Cross section of free-fall temperature
$T^2_{FFAR}$ for the $b=0.4, 0.5, 0.6, 1$ cases from top to down. The dashed
curve describes $T^2_{FFAR}$ vanishes at event horizon $x=1$.}
\label{fig.7}
\end{figure*}
\begin{figure*}[t!]
   \centering
   \includegraphics{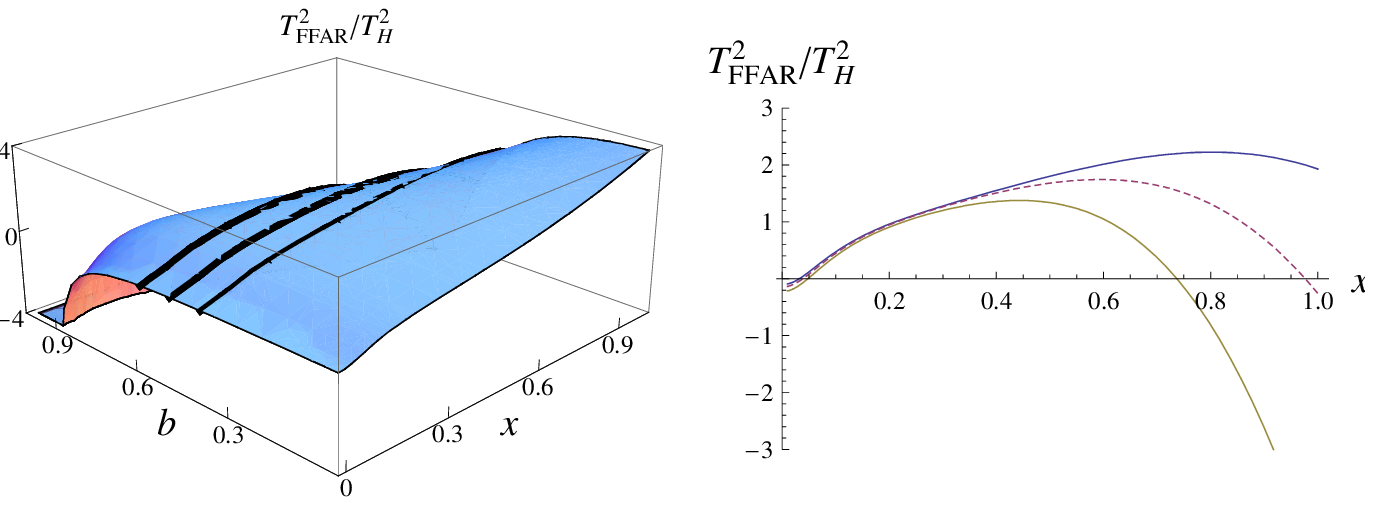}
\caption{Left panel: Free fall temperature $T^2_{FFAR}$ for the
RN-AdS black holes with $c=10$. The solid curves are for the $b=0.4,
0.5, 0.6$ cases. Right panel: Cross section of free fall temperature
$T^2_{FFAR}$ for the $b=0.4, 0.5, 0.6$ cases from top to down. The dashed
curve describes $T^2_{FFAR}$ vanishes at event horizon $x=1$.}
\label{fig.8}
\end{figure*}
On the other hand, Figs. \ref{fig.7} and \ref{fig.8} show the
squared free fall temperatures for the RN-AdS black holes
for fixed $c=100$, $c=10$ cases,
respectively, while for varying $b$. In Fig. \ref{fig.8}, the
curve of $b=1$ has been not shown because it has very large
negative value.

\begin{figure*}[t!]
   \centering
   \includegraphics{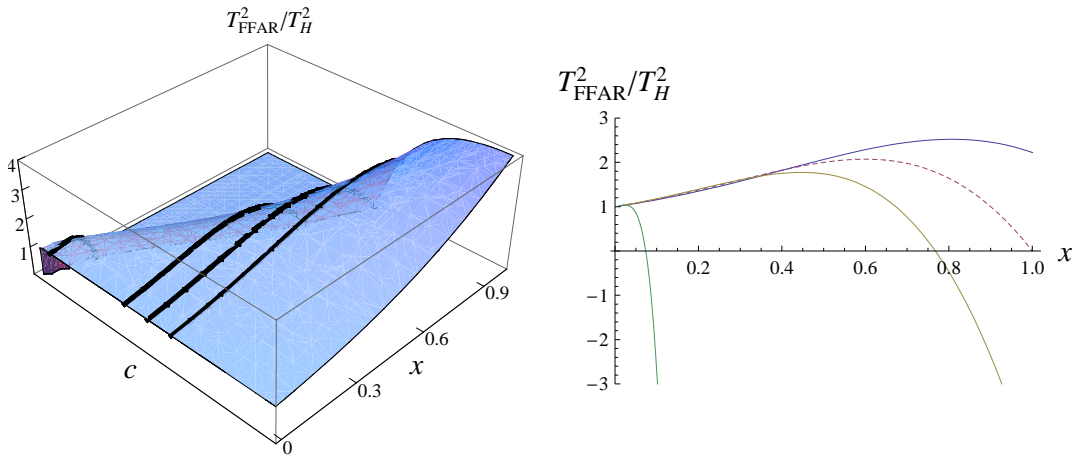}
\caption{Left panel: Free-fall temperature $T^2_{FFAR}$ for the
RN black holes ($c\rightarrow\infty$). The solid curves are for the
$b=0.4, 0.5, 0.6, 1$ cases. Right panel: Cross section of free-fall
temperatures $T^2_{FFAR}$ for the $b=0.4, 0.5, 0.6, 1$ cases from top to
bottom.} \label{fig.9}
\end{figure*}

Furthermore, Fig. \ref{fig.9} specializes to the $c\rightarrow\infty$ case,
the RN limit. At spacial infinity ($x\rightarrow 0$), the free
fall temperature coincides with the Hawking temperature
\begin{equation}
T_{FFAR}\rightarrow T_H=\frac{r_+-r_-}{4\pi r^2_+}.
\end{equation}
As a result, the difference between Fig. \ref{fig.7} ($c=100$) and
Fig. \ref{fig.9} ($c\rightarrow\infty$) lies in the behavior of
temperature in the asymptotically far away from the black holes: the
former $T^2_{FFAR}/T^2_H$ goes to zero, however, the latter goes to
one, which means that the free fall and Hawking temperatures
coincide.

\begin{figure*}[t!]
   \centering
   \includegraphics{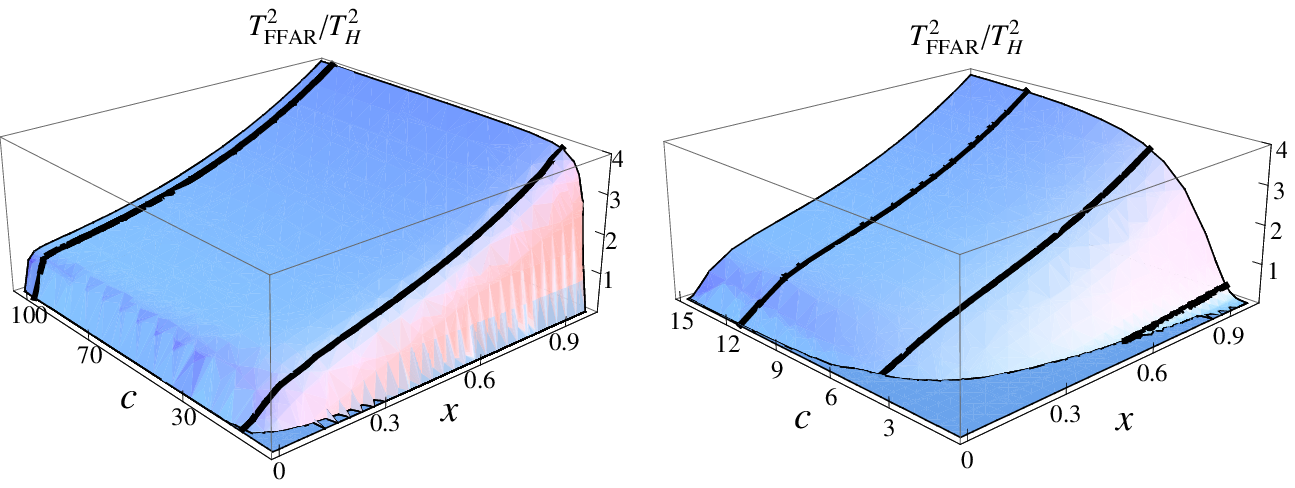}
\caption{Free-fall temperature $T^2_{FFAR}$ for $b=0$ in units of
$T^2_H$, which corresponds to the case of the SAdS black holes.
The solid curves are for the $c=100, 10, 5, 1$ cases.} \label{fig.10}
\end{figure*}

\begin{figure*}[t!]
   \centering
   \includegraphics{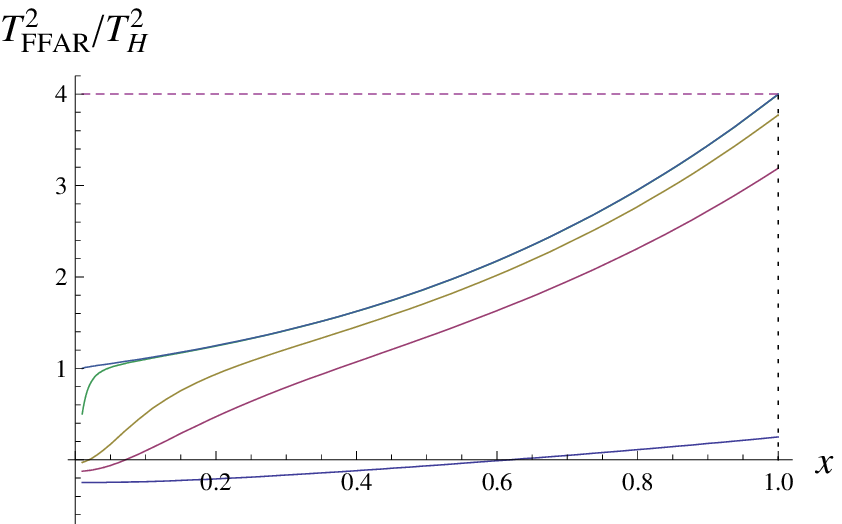}
\caption{Slices of free-fall temperature $T^2_{FFAR}$ for the $b=0$ case in
units of $T^2_H$. The curves are for the $c=1000, 100, 10, 5, 1$ cases from
top to down. } \label{fig.11}
\end{figure*}

On the other hand, Figs. \ref{fig.10} and \ref{fig.11} describe the SAdS limit of
$b\rightarrow 0$. Its asymptotic limit $x\rightarrow 1$ of the
local free temperature gives exactly Eq. (\ref{sadsT}), while on
event horizon one has
\begin{equation}
T^2_{FFAR}\rightarrow\frac{1}{4\pi^2r^2_+},
\end{equation}
as expected.
For a small AdS black hole with $r_H \ll l$ (or, large $c$), the
horizon area reduces to that of a RN black hole in asymptotically
flat spacetime with $r_H \approx m + \sqrt{m^2 - Q^2}$, and
(\ref{temp07}) reduces to
\begin{equation}
T_{FFAR} \rightarrow \frac{1}{2\pi r_H}\left(1 - 2b
\right)^{\frac{1}{2}} \approx 2T_H
\end{equation}
at the event horizon with $b \rightarrow 0~(Q^2 \ll r_H^2)$. This
can be seen in Fig. \ref{fig.11} near $x=1$ when $c=1.000$ (or,
$c=100$) and $b=0$. On the other hand, for a large AdS black hole
with $r_H \gg l$ (or, small $c$)~\cite{ht}, we find that at the
event horizon
\begin{equation}
T_{FFAR} \rightarrow \frac{1}{2\pi r_H} \ll T_H,
\end{equation}
which makes the ratio of $T^2_{FFAR}/T^2_H$ approximately zero.

Finally, we reobtain the Schwarzschild limit without the
cosmological constant from the SAdS embedding with the
$l\rightarrow \infty~(c\rightarrow \infty)$ limit, or the RN
embedding with the $Q = 0$~$(b=0)$ one. Taking the $c\rightarrow
\infty$ limit in Eq. (\ref{temp7s}), or the $b = 0$ limit in Eq.
(\ref{temp7r1}), the $T^2_{FFAR}$ and $T_H$ reduce to the
Schwarzschild one as follows
\begin{eqnarray} \label{temp7r}
 T^2_{FFAR} &=& \frac{1 + x + x^2+ x^3}
{16 \pi^2 r_H^2 },
\end{eqnarray}
which also coincide with the previous work \cite{bt}. As a final
comment, we have checked that the embedding functions (\ref{rna})
have exactly the same form of the ones in Ref. \cite{ksp,bt} when
taking the limit referred in this work.

\section{Summary}

In summary, we have used the global embedding of the RN-AdS black
hole spacetime into a (5+2) dimensional flat spacetime to define a
desired local temperature for observers in radial free fall outside a static
black hole as a generalization of the previous work \cite{bt}.
We have also shown that our extended results in the
GEMS of the RN-AdS space systematically include those of the
known limiting GEMS geometries, which are the RN,
SAdS, Schwarzschild, through the successive
truncation procedure of parameters in the original curved space.

\vskip 0.5cm

\section*{Acknowledgement}
The authors thank L. Thorlacius for helpful communication. Y.-W. Kim
was supported by the Korea Research Foundation Grant funded by Korea
Government (MOEHRD): KRF-2007-359-C00007. Y.-J. Park was supported
by the Korea Science and Engineering Foundation (KOSEF) grant funded
by the Korea government (MEST) (No. 20090083765 and R31-20002).


\end{document}